\documentclass[final,3p,times,twocolumn]{elsarticle}

\usepackage{amssymb}
\usepackage{amsmath}

\usepackage{hyperref}
\usepackage{color}
\usepackage{siunitx}
\usepackage{graphicx}
\usepackage{multirow}
\DeclareSIUnit \parsec {pc}

\journal{Astronomy and Computing}

\begin{document}

\begin{frontmatter}



\title{Optimising the FRB Search Pipeline for the Northern Cross Radio Telescope}


\author[um_label]{Hayley Camilleri} 
\author[um_label]{Alessio Magro}
\author[pavia,trento,inaf_cagliari]{Andrea Geminardi}
\author[inaf_bologna]{Giovanni Naldi}
\author[inaf_bologna]{Gianni Bernardi}
\author[inaf_bologna]{Luca Bruno}
\author[inaf_bologna]{Valentina Cesare}
\author[inaf_bologna]{Francesco Fiori}
\author[inaf_bologna]{Davide Pelliciari}
\author[inaf_cagliari]{Maura Pilia}
\author[inaf_cagliari]{Matteo Trudu}

\affiliation[um_label]{organization={Institute of Space Sciences and Astronomy (ISSA)},
            addressline={University of Malta}, 
            city={Msida},
            country={Malta}}
            
\affiliation[pavia]{organization={Scuola Universitaria Superiore IUSS of Pavia},
            city={Pavia},
            country={Italy}}

\affiliation[trento]{organization={University of Trento, Department of Physics},
                    city={Povo},
                    country={Italy}}

\affiliation[inaf_cagliari]{organization={Istituto Nazionale di Astrofisica (INAF), Osservatorio Astronomico di Cagliari},
            city={I-09047 Selargius (Cagliari)},
            country={Italy}}

\affiliation[inaf_bologna]{organization={Istituto Nazionale di Astrofisica (INAF), Istituto di Radio Astronomia},
            city={I-40129 Bologna},
            country={Italy}}

\begin{abstract}
Fast Radio Burst (FRB) search pipelines are being developed to operate under strict real-time constraints while maintaining sensitivity to short-duration transient signals. In incoherent dedispersion based pipelines such as Heimdall, apart from observation bandwidth and number of beams, detection performance and computational throughput are strongly dependent on the choice of processing parameters, which are often selected heuristically. In this work, we present a systematic evaluation of key dedispersion and matched filtering parameters and quantify their impact on both detection accuracy and runtime performance.\\

A controlled synthetic injection framework is developed in which artificial FRB pulses with known dispersion measures (DMs), signal-to-noise ratios (SNRs), and pulse widths are embedded into realistic filterbank data containing instrumental noise representative of observations from the Northern Cross radio telescope. Using this framework, a grid of Heimdall configurations is explored, spanning DM tolerance, boxcar filter width, and processing gulp size. Detection performance is assessed by comparing recovered and injected signal properties, while computational performance is evaluated through end-to-end processing time measurements.\\

The results reveal clear trade-offs between sensitivity and throughput across parameter choices. We identify an empirically optimal configuration that provides burst recovery while maintaining processing speeds exceeding real-time requirements. While the specific optimal parameters are derived for the Northern Cross, the methodology and findings are broadly applicable to any real-time transient detection pipeline employing matched-filtering and dedispersion, and are particularly relevant for low-frequency radio telescopes with similar observing configurations. These findings demonstrate the value of data-driven parameter evaluation for improving the performance of real-time transient detection pipelines.
\end{abstract}




\end{frontmatter}



\section{Introduction}
\label{sec1}

Fast Radio Bursts (FRBs) are bright, millisecond-duration radio transients, first identified as a distinct phenomenon through the discovery of a highly dispersed burst in archival pulsar survey data \cite{lorimer07} and subsequently established as a population through multiple detections at cosmological distances \cite{thornton13}. Their dispersion measures (DMs) frequently exceed the expected Galactic contribution, indicating an extragalactic origin and enabling the use of FRBs as probes of ionised baryons along the line of sight \cite{cordes03}. Beyond their utility as cosmological and intergalactic medium probes, FRBs exhibit a wide diversity in temporal and spectral structure, including complex sub-burst morphology, strong polarisation, and in some cases multiple detections, motivating extensive observational campaigns and rapid follow-up strategies across the electromagnetic spectrum \cite{petroff19}.\\

Within this landscape, the Northern Cross radio telescope represents a compelling case study for pipeline development and optimisation \citep{perley19}. The Northern Cross operates in the 400-\SI{416}{\mega \hertz} band and has a large collecting area and wide field of view, making it well suited to FRB searches at low frequencies \cite{locatelli20,debarro25}. At these frequencies, dispersion delays across the band are substantial for high-DM events (a DM $=500$ event would spread across the band for about 1\si{\second}), and pulse broadening effects can be significant, increasing the importance of carefully tuned dedispersion and matched-filtering settings. Currently, the instrument is undergoing a major refurbishment and digital upgrade, including the deployment of a modern infrastructure for acquisition and processing and the installation of High-Performance Computing (HPC) resources designed to support full-instrument FRB searches in real time \cite{debarro25,naldi25}. This upgrade motivates the need for principled, data-driven evaluation of transient search pipeline configurations, ensuring that the upgraded system can achieve reliable burst recovery while meeting real-time throughput constraints.\\

Heimdall is a widely used GPU-accelerated single-pulse search tool that implements brute-force incoherent dedispersion and boxcar-based matched filtering \cite{barsdell12}. It is used in several transient search backends due to its flexibility and suitability for real-time processing \cite{keane18survey, keane18}. However, Heimdall’s performance is strongly influenced by user-defined parameters such as DM tolerance and boxcar filter widths. In practice, these parameters are often selected heuristically, or inherited from previous surveys, with limited quantitative assessment of their impact on detection accuracy or computational efficiency.\\

The lack of systematic evaluation of pipeline parameter choices presents a challenge for real-time transient searches, particularly for telescopes operating under hardware or latency constraints. Suboptimal configurations may lead to unnecessary computational overhead, reduced sensitivity to specific classes of bursts, or increased false-positive rates. Previous studies have examined search sensitivities and aspects of algorithmic performance in FRB surveys \cite{keane15,qiu23}. Further work has also explored alternative dedispersion strategies and candidate classification methods, including machine-learning-based approaches \cite{connor16,agarwal20,connor18}. However, relatively little attention has been devoted to comprehensive, multi-metric evaluation of parameter-level trade-offs within established incoherent dedispersion pipelines, particularly in the context of jointly optimising detection fidelity and real-time computational performance for specific instrument configurations.\\

In this work, we present a systematic, data-driven evaluation of key processing parameters in an incoherent dedispersion-based FRB search pipeline using Heimdall for the Northern Cross telescope. We employ a controlled synthetic injection framework in which artificial FRB signals with known properties are embedded into true filterbank data. This approach enables direct, quantitative comparison between injected and recovered burst properties, allowing both detection accuracy and runtime performance to be assessed across a grid of parameter configurations.\\

The goals of this study are twofold: first, to characterise the trade-offs between sensitivity and computational throughput associated with commonly used Heimdall parameters; and second, to identify empirically optimal configurations that satisfy real-time processing requirements while maintaining transient recovery. Although the experiments are motivated by the Northern Cross telescope, the methodology and conclusions are broadly applicable to incoherent dedispersion pipelines used in contemporary radio transient searches. The results presented here should therefore be interpreted as both instrument-specific recommendations and a general demonstration of how systematic parameter evaluation can inform pipeline optimisation for incoherent dedispersion based transient searches.

The remainder of this paper is organised as follows. Section \ref{sec:background} provides theoretical background on signal dispersion and propagation effects relevant to FRB detection, along with an overview of contemporary real-time FRB detection systems. Section \ref{sec:telescope} describes the Northern Cross radio telescope and its observing configuration. Section \ref{sec:heimdall} introduces the Heimdall pipeline and defines the parameter space explored in this study, covering the dedispersion and matched-filtering framework, the specific parameter combinations evaluated, and the computational environment used. Section \ref{sec:injection} presents the synthetic injection framework and evaluation methodology, including the signal generation procedure, detection accuracy and runtime metrics, and the statistical analysis approach comprising dimensionality reduction, unsupervised clustering, and non-parametric hypothesis testing. Section \ref{sec:results} reports the results across the full parameter grid, covering detection accuracy, runtime scaling, statistical comparisons, identification of the empirically optimal configuration, and the cluster structure of the performance space. Section \ref{sec:discussion} discusses the broader implications of the findings, and Section \ref{sec:conclusion} summarises the conclusions and outlines directions for future work.

\section{Theoretical Background}
\label{sec:background}

\subsection{Dispersion and Propagation Effects}

The presence of free electrons in the ISM and intergalactic medium causes signal dispersion, which is a distinguishing property of FRBs. The frequency dependent delay caused by this plasma dispersion is a determining characteristic used in detection. The temporal delay between two frequencies is determined by the equation,

\begin{equation}
    \Delta t = 4.15 \times 10^3 \mathrm{ms} \left( \frac{1}{\nu_1^2} - \frac{1}{\nu_2^2} \right) \times \left(\frac{\text{DM}}{\mathrm{pc\,cm} ^{-3}}\right),
\end{equation}

where $\nu_1$ and $\nu_2$ are frequencies in \unit{\mega \Hz} and $\text{DM}$ is the integrated column density of free electrons along the line of sight \citep{lorimer24}. The dispersion measure itself is given by,

\begin{equation}
    \text{DM} = \int_{0}^{d} n_e (l) \, dl,
\end{equation}

where $n_e(l)$ is the electron density at a distance $l$ and $d$ represents the distance from the Earth to the pulsar \citep{draine11}.\\

\begin{figure}[h]
    \centering
    \hbox{\hspace{-1.5em}\includegraphics[width=1.1\linewidth]{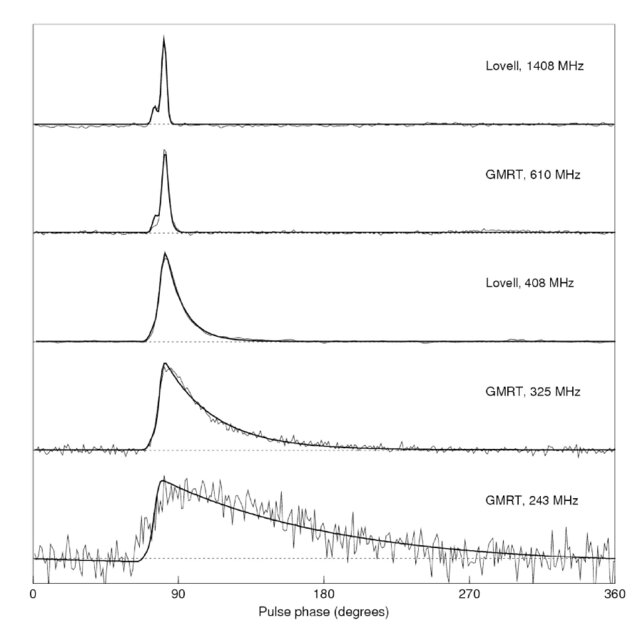}}
    \caption{Pulse profiles for PSR B1831-03 observed at five different frequencies with the Lovell telescope and the GMRT, clearly showing the increasing effect of scattering at lower frequencies. The solid lines show exponential fits to the data. Figure extracted from \cite{lorimer05}.}
    \label{fig:scatt}
\end{figure}

In addition to dispersion, FRBs often show propagation effects such as scattering, which is depicted in Figure \ref{fig:scatt}, resulting in asymmetric pulse broadening, especially at lower frequencies. The scattering measure is the path integral of $C_n^2$ \citep{cordes03},

\begin{equation}
    \text{SM} = \int_0^D ds \, C_n^2
\end{equation}

where $D$ represents the independent distance measurements and $C_n^2$ is the spectral coefficient (the “level of turbulence”).\\

Scintillation, generated by small scale turbulence in the plasma, introduces variation in signal intensity. Faraday rotation, which rotates the polarisation angle with frequency, offers information on the magnetic field intensity and electron density of the intervening medium. Some bursts exhibit indications of plasma lensing, which occurs when inhomogeneities in the plasma cause the signal to focus or defocus. These effects limit detection while providing a lot of information about the cosmic environment \citep{cordes16a}.\\

\subsection{FRB Detection Systems}

Real-time detection has become central to FRBs because it enables prompt alerts and triggered capture of high-time-resolution data, which are essential for studying burst microstructure, polarisation, and for achieving improved localisation. In addition, real-time processing is decisive for managing the large data volumes produced by modern radio telescopes operating at high time and frequency resolution, allowing transient events to be identified and retained while reducing the need to store and process prohibitively large raw data streams. As summarised in recent pipeline focused work \cite{geminardi25, pelliciari24}, multiple observatories have deployed low-latency FRB detection systems with varying architectures, including image-plane or beamformed searches and voltage-buffer triggering \cite{debarro25}. Representative examples include the VLA realfast system, which performs commensal transient searching with rapid processing of interferometric data \cite{realfast18}, as well as ASKAP’s CRAFT program, which conducts commensal real-time searches and supports voltage capture for localisation and high-time-resolution studies \cite{craftics}. Complementary approaches have been demonstrated at other facilities, including UTMOST real-time detections with voltage capture \cite{utmost19} and real-time FRB searching systems developed for FAST \cite{fastsys23}. Collectively, these systems illustrate the state of the art: modern FRB surveys increasingly rely on GPU-accelerated pipelines capable of sustained high-throughput processing, low-latency candidate generation, and robust event triggering.\\

Over the past decade, the field has transitioned from isolated discoveries to systematic surveys that detect FRBs at high rates and publish large, uniform samples. A major milestone was the release of the first CHIME/FRB catalog, comprising of 536 FRBs detected between 400–\SI{800}{\mega \hertz} in a uniform survey with calibrated selection effects \cite{chimecat1}. Follow-up analyses have leveraged channelised raw voltage data for subsets of events to refine burst properties and enable higher-fidelity characterisation \cite{chimeupdate24}. More recently, large-sample catalogues have expanded dramatically in size, enabling population-level studies of repetition, energetics, and selection biases with unprecedented statistical power \cite{chimecat2}. These developments have established FRBs as a mature time-domain field in which discovery rates and scientific yield are increasingly limited not by telescope sensitivity alone, but by the capability of real-time processing systems to detect, classify, and trigger on events reliably.\\

\section{The Northern Cross Radio Telescope}
\label{sec:telescope}

\begin{figure}
    \centering
    \includegraphics[width=\linewidth]{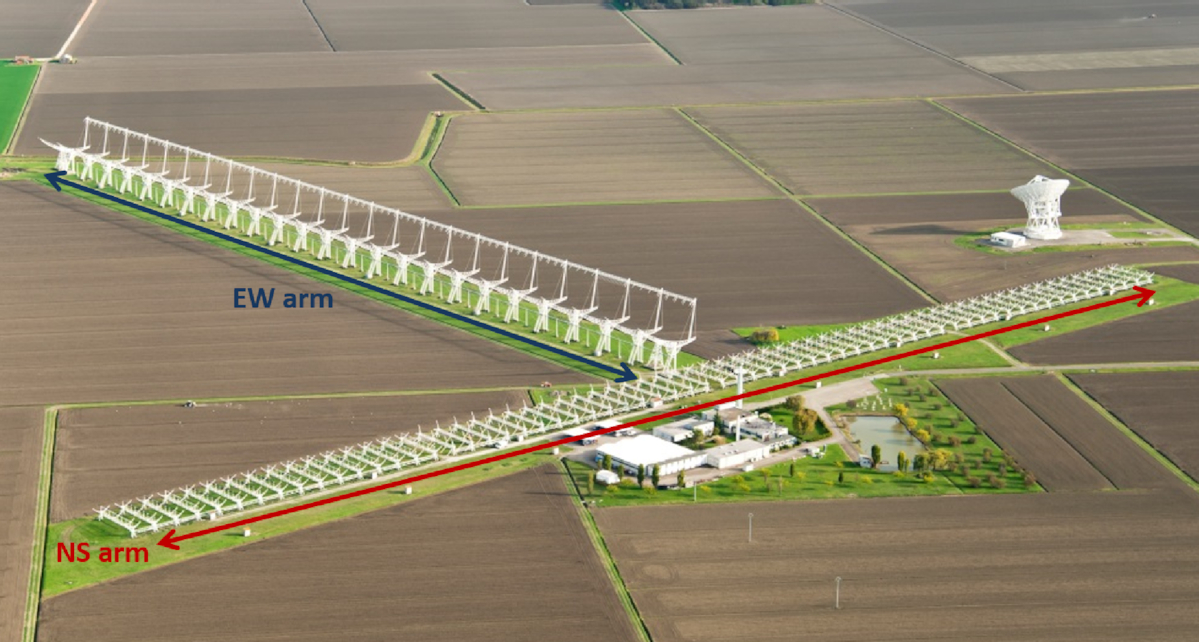}
    \caption{Top view of the Northern Cross radio telescope with
the two perpendicular arms along the East-West and North-South
direction. Figure extracted from \cite{debarro25}.}
    \label{fig:NSWE_NC}
\end{figure}

The experiments presented in this work are motivated by observations from the Northern Cross radio telescope, a transit radio interferometer located near Bologna, Italy (See Figure \ref{fig:NSWE_NC}). Originally designed for wide-area radio surveys, the Northern Cross is now being repositioned as a competitive instrument for time-domain astronomy, with a particular focus on FRB detection at low radio frequencies.\\

The telescope operates at a central observing frequency of approximately 408 MHz with a bandwidth of 16 \unit{\mega \Hz}, typically channelised into 1024 frequency channels \citep{locatelli20}. It should be noted that these parameters are subject to change following future instrument upgrades \citep{pilia20}, which would directly affect the dispersive smearing timescale and consequently the optimal DM trial spacing and parameter selection discussed in this work. This fine spectral resolution enables accurate tracking of dispersion delays across the band and facilitates the detection of highly dispersed transient signals \cite{debarro25}. At these frequencies, dispersion delays are significant for highly dispersed extragalactic FRBs, though not extreme in absolute terms given the modest bandwidth. The primary computational challenge instead arises from the high time and spectral resolution of the backend, which produces large data rates and millions of time samples per minute of observation. Performing brute-force incoherent dedispersion across wide DM ranges under these conditions places stringent demands on GPU throughput and memory bandwidth in real-time operation \cite{cordes03}. Consequently, the Northern Cross provides a representative and challenging test case for evaluating incoherent dedispersion based FRB search pipelines.\\

As described by \cite{debarro25}, the recent upgrade of the Northern Cross includes a new digital acquisition and processing chain designed to support real-time FRB searches, with GPU-accelerated pipelines enabling low-latency transient detection across wide DM ranges \cite{barsdell12,debarro25}. At present, observations are stored as filterbank files and analysed offline using software such as Heimdall, providing a convenient framework for controlled performance evaluation prior to full real-time deployment. A key challenge of the full-instrument configuration is that all simultaneously formed beams must be processed in parallel, generating substantial data rates and computational loads that scale rapidly with the number of beams, DM trials, and time samples. In this context, conservative or poorly tuned parameter choices can lead to unnecessary overhead, reduced sensitivity, or failure to meet real-time processing requirements.\\

The transition from a legacy backend to a modern real-time FRB search system therefore motivates the need for systematic, quantitative evaluation of pipeline parameter configurations. Rather than adopting parameter settings by analogy with other telescopes or surveys operating at different frequencies and bandwidths, the upgraded Northern Cross requires tuning that explicitly accounts for its observing band, dispersion regime, and available computational resources. The work presented here addresses this requirement by using controlled synthetic injections to evaluate the impact of dedispersion granularity, matched-filter coverage, and buffering strategy on both detection fidelity and runtime performance.\\

Data acquired by the Northern Cross are recorded as frequency-time filterbank files with fixed time and frequency resolution, typically $\sim$\SI{80}{\micro\second} time sampling and $\sim$\SI{200}{\kilo\Hz} channel widths across the 400-416\si{\mega\Hz} band. This high time and spectral resolution preserves sensitivity to narrow, highly dispersed bursts, but substantially increases data volume and computational cost for real-time dedispersion and matched filtering. These data products are compatible with standard transient search software, including Heimdall, and retain the instrumental noise and system characteristics present in real observations. While the telescope’s observing strategy and backend architecture impose specific constraints on data rates and processing latency, the core signal processing challenges, dedispersion across wide DM ranges and matched filtering for short-duration pulses, are common to many contemporary FRB search pipelines.

\section{Heimdall Pipeline and Parameter Space}
\label{sec:heimdall}

This section describes the FRB search evaluated in this study and outlines the key processing parameters explored. We first summarise the core signal-processing stages implemented by the Heimdall single-pulse search software, focusing on incoherent dedispersion and matched filtering for transient detection. We then define the parameter space considered in this work, highlighting how choices related to DM sampling, filter widths, and data buffering directly influence both detection accuracy and computational performance. Together, these elements establish the framework within which the systematic evaluation presented in subsequent sections is conducted.

\subsection{Incoherent Dedispersion and Single Pulse Search}

The FRB search evaluated in this work is through the use of Heimdall, which performs brute-force dedispersion followed by matched filtering to identify transient signals \cite{barsdell12}. Input data are provided as frequency-time filterbank files, which are dedispersed across a user-defined grid of DMs. For each trial DM, Heimdall applies frequency-dependent time shifts to correct for dispersion delays introduced by the ionised interstellar medium, producing a one-dimensional dedispersed time series \cite{cordes03}.\\

\begin{figure}[h]
    \centering
    \hbox{\hspace{-1.5em}\includegraphics[width=1.1\linewidth]{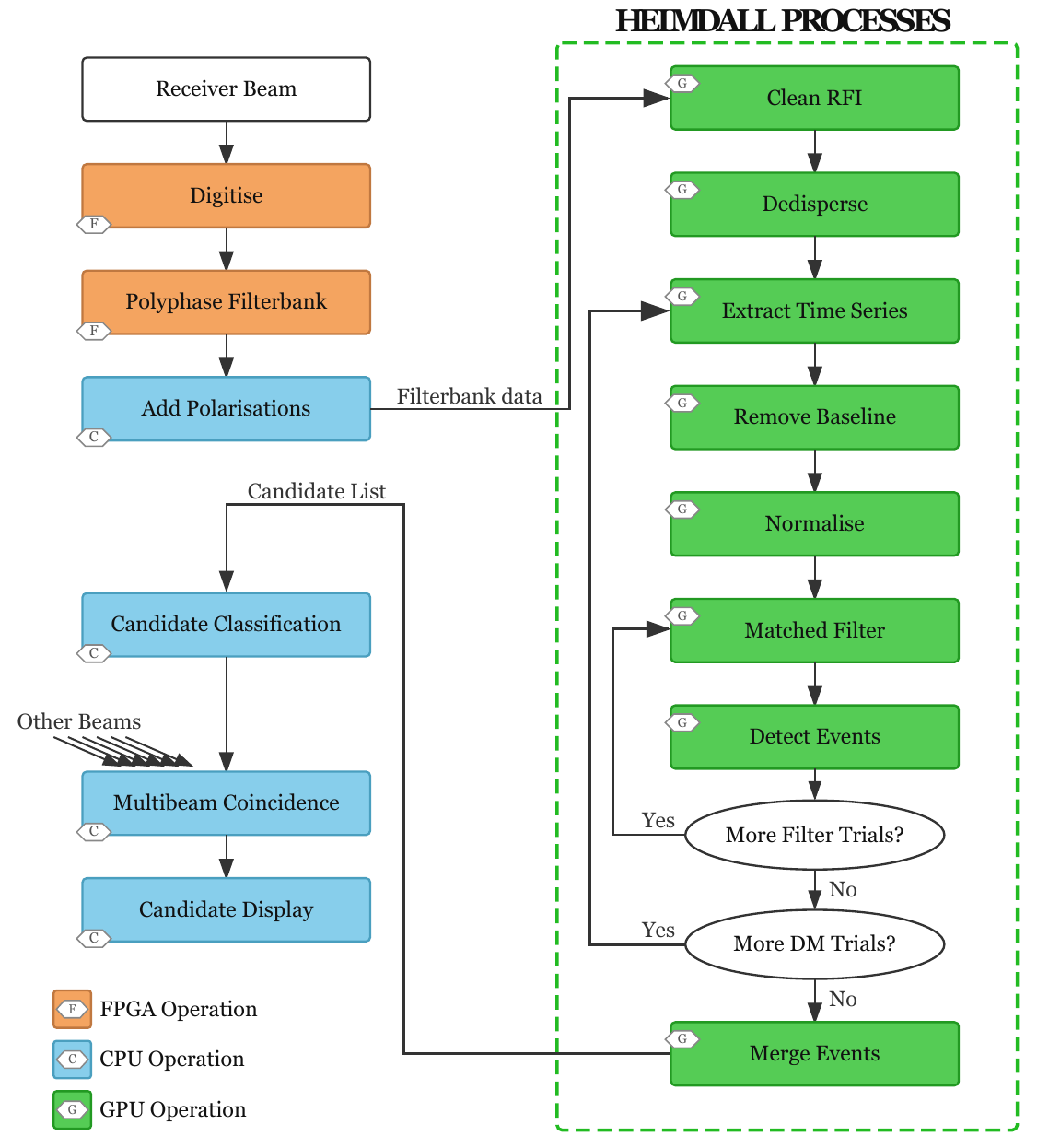}}
    \caption{Flow chart of the key processing operations in the pipeline. Heimdall is the name of the main GPU-based pipeline implementation. Adapted from \cite{barsdell12}.}
    \label{fig:heimdall}
\end{figure}

Following dedispersion, Heimdall performs a single-pulse search by convolving each dedispersed time series with a set of boxcar filters of increasing width. These boxcar filters act as matched filters for pulses of varying temporal extent, enhancing the SNR when the filter width approximately matches the intrinsic pulse width \cite{cordes04}. Candidate events are identified as statistically significant peaks in the filtered time series that exceed a predefined SNR threshold. For each detected candidate, Heimdall records properties including Time of Arrival (ToA), DM, pulse width, and SNR.\\

This brute-force approach is computationally intensive but offers flexibility and robustness across a wide range of pulse morphologies and DMs. However, the overall performance of the pipeline is dependent on the configuration of several user-defined parameters that control the granularity of the dedispersion, the filtering strategy, and the buffering of the data. A flow chart showing all the steps that are computed by Heimdall, as well as additional ones that are performed in an FRB search pipeline, can be found in Figure \ref{fig:heimdall}.

\subsection{Parameter Space Explored}

This study focuses on three key Heimdall parameters that strongly influence both detection performance and computational cost: DM tolerance, boxcar filter width, and gulp size.\\

The DM tolerance parameter (\texttt{dm\_tol}) controls the spacing of trial DMs and effectively determines the maximum allowable fractional SNR loss due to dedispersion mismatch between adjacent DM trials \cite{barsdell12}. This parameter governs the adaptive spacing of the DM trial grid; each trial is placed such that the effective pulse width grows by a factor of \texttt{dm\_tol} from one trial to the next. The effective pulse width at any given DM trial is given by:

\begin{equation}
    W_{eff} = \sqrt{t_{int}^2 + t_{samp}^2 + t_{DM}^2 + t_{\delta DM}^2 + \tau_s^2}
\end{equation}

where $t_{int}$ is the intrinsic pulse width, $t_{samp}$ is the sampling time, $t_{DM}$ is the dispersive smearing across a single frequency channel, $t_{\delta DM}$ is the smearing introduced by the offset between the true DM and the nearest trial DM, and $\tau_s$ is the scattering timescale. Lower DM tolerance values result in finer DM grids and improved sensitivity at the cost of increased computational load, while higher values reduce the number of trials but may degrade pulse recovery for signals whose true DM lies between grid points. It is worth noting that \cite{keane26} highlight that the relationship between \texttt{dm\_tol} and actual survey sensitivity is not entirely straightforward due to the scalloped response between trials, meaning the true worst case SNR loss is not simply $1/$\texttt{dm\_tol}.\\

Boxcar filtering is performed using a predefined set of filter widths, log$_2$ spaced, expressed in samples. Wider boxcars improve sensitivity to broader pulses but increase computational complexity and susceptibility to noise integration. Conversely, narrow boxcars favour short duration pulses but may underperform for temporally broadened signals. In this work, we explore multiple boxcar configurations to assess how the upper bound of the filter width range affects detection accuracy and runtime.\\

The gulp size parameter specifies the duration of data processed in each iteration. Larger gulp sizes can improve GPU utilisation and reduce kernel launch overhead, but they also increase memory requirements and may introduce additional latency. Smaller gulp sizes reduce buffering latency but may lead to suboptimal throughput. We therefore investigate the effect of varying gulp size on real-time processing performance.\\

\begin{table}[h]
    \centering
    \begin{tabular}{cccccc}
        \hline
        DM Tolerance & 1.001 & 1.01 & 1.05 & 1.1 & 1.2 \\
        Boxcar Width & 32 & 64 & 128 & 256 & 512 \\
        \hline
    \end{tabular}
    \caption{Parameter Combination Values.}
    \label{tab:parameters}
\end{table}

A grid of parameter combinations was constructed by varying DM tolerance and maximum boxcar width across ranges representative of practical FRB search configurations, which can be seen in Table \ref{tab:parameters}. This parameter space was chosen to reflect both commonly used settings and more aggressive configurations that trade sensitivity for computational efficiency.

\subsection{Computational Environment}

All experiments were conducted on a GPU-accelerated computing system representative of the operational environment used for FRB searches at the Northern Cross radio telescope. The pipeline was executed on a single NVIDIA RTX 6000 Ada GPU, and runtime measurements were obtained using end-to-end processing times reported by Heimdall. Heimdall \footnote{\url{https://sourceforge.net/p/heimdall-astro/wiki/Home/}} was used in its standard, unmodified form, ensuring that all observed performance differences arise solely from pipeline parameter selection rather than changes to the underlying implementation.\\

While the numerical values of optimal parameters may depend on specific hardware characteristics, the relative performance trends and trade-offs identified in this study are expected to be broadly applicable to similar GPU-based incoherent dedispersion pipelines.

\section{Synthetic Injection Framework and Evaluation Metrics}
\label{sec:injection}

\subsection{Synthetic FRB Signal Generation}

The signals have been injected in real observations from the Northern Cross radio telescope. The sample of filterbanks has been selected in a way that we do not expect real astrophysical signals inside. Indeed, we used Northern Cross observations of extragalactic FRBs already analysed with tested pipelines, which excluded the presence of radio bursts, and at high sky declination to avoid the signals coming from the Galactic plane. No RFI cleaning tool has been used in order to simulate real observations.\\

To enable controlled and reproducible evaluation of pipeline performance, we employ a synthetic signal injection framework in which artificial FRB-like pulses with known properties are embedded into filterbank data using the published package \texttt{FRB Faker} \cite{houben25}. Synthetic injections provide direct ground truth, allowing quantitative assessment of detection accuracy and computational performance without the ambiguities inherent in real observational data \cite{petroff19}.\\

Each synthetic burst is generated with a Gaussian temporal profile and injected into the dynamic spectrum prior to dedispersion. The dispersion delay across frequency channels is applied using the cold plasma dispersion relation \cite{cordes03}, ensuring that injected signals exhibit realistic frequency-dependent arrival times. DMs are sampled from a log-uniform distribution spanning 20 to 3000 \si{\parsec \per \centi \metre \cubed}, reflecting the wide dynamic range of DMs observed in FRB populations while avoiding over-representation of low-DM events.\\

The injected pulses were generated with input SNR values sampled uniformly from the range [3, 13]. It is important to note that the input SNR defined by \texttt{FRB Faker} does not correspond directly to the SNR reported by Heimdall, as SNR estimation in radio filterbank data is not standardised and varies with pulse width, observing parameters, and the detection algorithm employed. As noted by the \texttt{FRB Faker} developers, a scaling coefficient is generally required to map between the two definitions. Given this ambiguity, and following the recommendation of the dataset authors, the SNR values used throughout this analysis are those reported directly by Heimdall, as these are the quantities directly relevant to the detection pipeline under evaluation. The near complete recovery of all injected pulses across all settings is consistent with the effective Heimdall SNR of all injections exceeding the pipeline detection threshold, despite some injections having low input SNR values. This reflects the empirical nature of the input SNR range, which was chosen to produce a representative distribution of weak and strong bursts in the filterbank rather than to correspond to specific Heimdall detection thresholds.\\

Pulse widths are drawn from a uniform distribution between 0.5 \si{\milli \second} and 130 \si{\milli \second}, covering both narrow, unresolved pulses and broader events potentially affected by temporal scattering, where multi-path propagation through turbulent plasma broadens the signal and produces an asymmetric pulse profile with an extended exponential tail. While real FRBs often exhibit more complex temporal and spectral structure, including asymmetric scattering tails and sub-burst components, the use of Gaussian pulses provides a consistent and interpretable basis for comparative evaluation of pipeline parameters.\\

Each filterbank file contains multiple injected bursts at random arrival times, ensuring that the pipeline is evaluated across a diverse set of signal properties and temporal contexts. The underlying data preserve realistic noise characteristics, channelisation, and time resolution, ensuring that injected signals are evaluated under conditions comparable to operational FRB searches.\\

Injection is performed prior to dedispersion and single pulse searching, allowing the full pipeline including dedispersion, boxcar filtering, and candidate selection to operate on the modified data without additional intervention. This approach ensures that recovered signal properties can be directly compared to known injection parameters, and that runtime measurements reflect end-to-end pipeline behaviour. A total of 950 filterbank files of 140 seconds of duration are generated, each containing 13 injected bursts, resulting in a dataset comprising 12,350 synthetic FRB events.

\subsection{Evaluation Metrics}
\label{eval_metrics}

Performance is evaluated using a combination of detection accuracy and computational efficiency metrics. Detection accuracy is assessed by matching detected candidates to injected bursts based on temporal proximity and DM consistency. For matched events, we compute relative errors between injected and recovered values of DM, arrival time, pulse width, and SNR.\\

\begin{figure*}
    \centering
    \includegraphics[width=\linewidth]{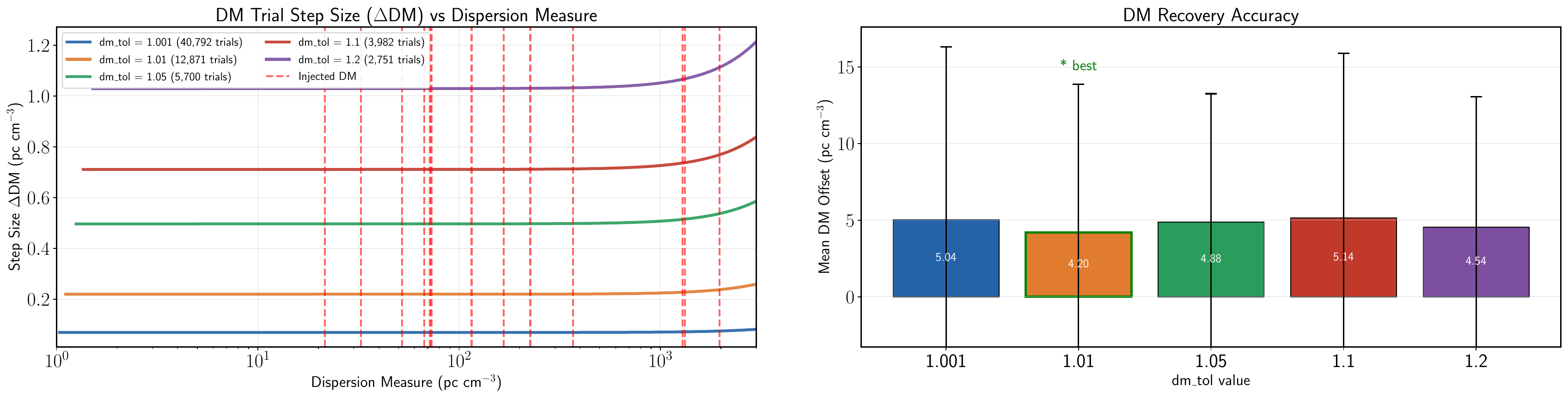}
    \caption{Impact of the \texttt{dm\_tol} parameter on DM trial spacing and recovery accuracy for Heimdall dedispersion. The left panel shows the local DM trial step size ($\Delta$DM) as a function of dispersion measure, with red dashed lines marking the injected pulse DM positions. The right panel shows the mean absolute offset between detected and injected DM values $(\pm 1\sigma)$ for each \texttt{dm\_tol} setting, where all values successfully recover injections but with varying accuracy, demonstrating that parameter choice affects signal recovery fidelity rather than detection alone. Results shown are from a representative file drawn from the test dataset; the trends are consistent across the full dataset.}
    \label{fig:dm_rec}
\end{figure*}

Rather than adopting a binary detection metric, this approach enables a nuanced assessment of how parameter choices affect the fidelity with which signal properties are recovered. This is particularly important for evaluating dedispersion and matched-filtering performance, where incorrect parameter settings may still produce detections but with degraded accuracy. As illustrated in Figure \ref{fig:dm_rec}, the \texttt{dm\_tol} parameter directly controls the spacing between DM trials ($\Delta$DM): larger values produce coarser trial grids while requiring fewer total trials, whereas smaller values sample the DM space more finely at greater computational cost (upper panel). Noticably, all \texttt{dm\_tol} values tested successfully recovered all injected pulses across the DM range, demonstrating that detections occur regardless of the parameter setting. However, the mean absolute DM offset between the detected and injected DM values, which is a direct measure of recovery accuracy, varies significantly across settings, with \texttt{dm\_tol} = 1.1 also exhibiting the largest spread in recovery accuracy (lower panel). This confirms that while all settings produce detections, the fidelity with which the DM, and by extension other derived signal properties, is recovered highly depends on the choice of \texttt{dm\_tol}, motivating a careful parameter selection rather than reliance on default values.\\

\subsection{Runtime Performance}
\label{run_performance}

Computational efficiency is assessed using the total processing time required to analyse each filterbank file. Runtime measurements are obtained directly from Heimdall’s internal timing reports and include dedispersion, boxcar filtering, and candidate generation. Input/output operations are excluded to ensure that runtime comparisons reflect pipeline performance rather than storage system characteristics.\\

Runtime is reported both in absolute terms and relative to the duration of the input data, allowing direct assessment of real-time feasibility. Configurations achieving processing speeds exceeding real-time requirements are considered operationally viable, while slower configurations are deemed unsuitable for real-time deployment despite potential gains in sensitivity.

\subsection{Statistical Analysis}

To analyse the multi-dimensional space arising from the evaluated parameter configurations, we employ a combination of exploratory visualisation, unsupervised clustering, and non-parametric statistical testing. This layered approach allows qualitative structure in the data to be identified prior to formal hypothesis testing, and ensures that statistically significant differences are interpreted in the context of overall performance trends.\\

\subsubsection{Dimensionality Reduction with t-SNE}

\begin{figure}[h!]
    \centering
  {%
    \includegraphics[width=\linewidth]{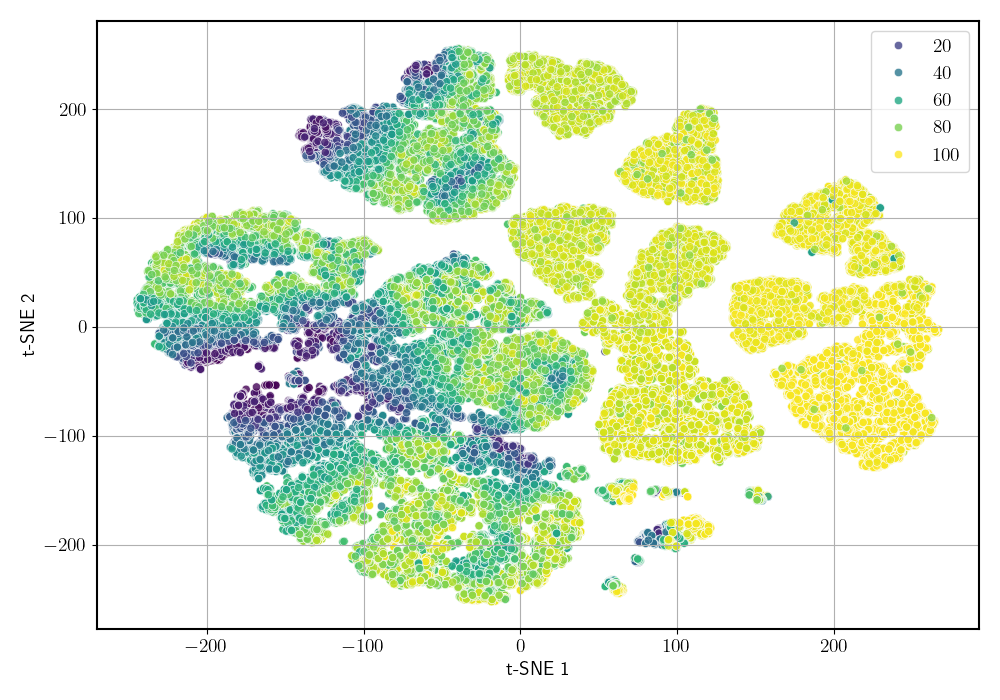}}%
  \hfill
  {%
    \includegraphics[width=\linewidth]{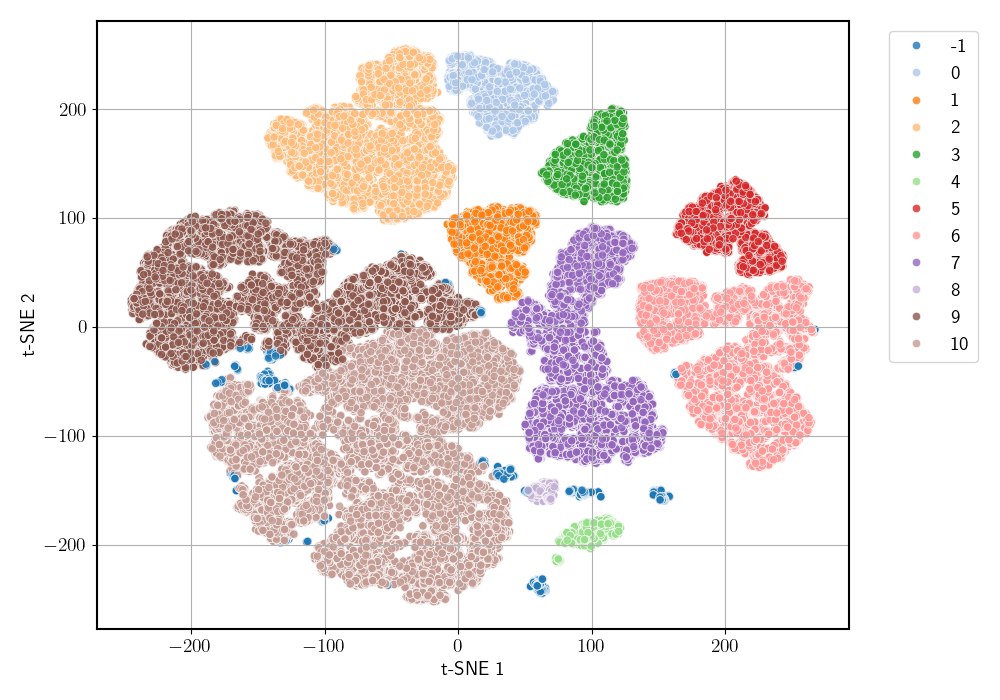}}%
        
    \caption{2-D projection of t-SNE dimensionality reduction on the data; where the top figure is overlayed with a heatmap representing percentage accuracy for SNR (cool colours = lower accuracy; warm colours = higher accuracy, up to 100\%) and the bottom figure is overlayed with colours which indicate cluster labels returned by HDBSCAN; label –1 marks points classified as noise/outliers. Each point represents a parameter-file outcome embedded into two dimensions by t-SNE (axes are unitless and not directly interpretable). High-density regions indicate many outcomes with very similar feature profiles (locally preserved neighbourhoods), i.e., parameter combinations that obtained similar performance and results.}
    \label{fig:t-sne_projections}
\end{figure}


As an initial exploratory step, we apply t-distributed stochastic neighbour embedding (t-SNE) to project the high-dimensional performance metrics into a two-dimensional space for visual analysis \cite{maaten08} as represented in Figure \ref{fig:t-sne_projections}. The input feature space includes detection accuracy metrics (DM error, SNR error, and ToA error) together with runtime performance, enabling joint assessment of sensitivity and computational efficiency. Prior to dimensionality reduction, all features are standardised to ensure comparable scaling and to prevent dominance by any single metric.\\


t-SNE is a non-linear dimensionality reduction technique that seeks to preserve local neighbourhood structure when mapping data from a high-dimensional space into a lower-dimensional embedding. In the original feature space, t-SNE models pairwise similarities between points using conditional probability distributions derived from Gaussian kernels, with the kernel bandwidth determined by a user-defined perplexity parameter. In the low-dimensional embedding, similarities are modelled using a heavy-tailed Student t-distribution, which reduces the crowding problem and allows moderately distant points to be more effectively separated.\\

The optimisation objective of t-SNE minimises the Kullback-Leibler divergence between the high-dimensional and low-dimensional similarity distributions. As a result, points that are close neighbours in the original feature space are encouraged to remain close in the embedding, while large pairwise distances are not preserved in a metric sense. For this reason, t-SNE embeddings should not be interpreted as preserving global geometry or absolute distances, but rather as providing a faithful representation of local relationships among configurations.\\

In this study, t-SNE is used solely as a visualisation tool, allowing intuitive inspection of whether parameter configurations form natural groupings or exhibit trade-off structures in the combined accuracy–runtime space. As an initial step to identify certain performance trends in the t-SNE projection, a heatmap was used to overlay the data points which represents the accuracy scores of the SNR feature, as can be seen in Figure \ref{fig:t-sne_projections}. This visualisation illustrates regions with changing signal clarity, finding potential clusters associated with greater or lower SNR accuracy values.

\subsubsection{Unsupervised Clustering with HDBSCAN}
\label{cluster}

To objectively identify groups of parameter configurations with similar performance, we apply the Hierarchical Density-Based Spatial Clustering of Applications with Noise (HDBSCAN) algorithm to the reduced performance space \cite{campello13}. HDBSCAN extends the DBSCAN framework by constructing a hierarchy of density-based clusters and extracting the most stable groupings.\\

Unlike partition-based clustering methods such as k-means, HDBSCAN does not require specification of the number of clusters and is capable of identifying clusters of varying density while explicitly labelling outliers as noise. This is particularly advantageous in the present context, where performance distributions are heterogeneous and some parameter configurations may represent suboptimal or extreme cases rather than belonging to well-defined groups.\\


The clustering results, depicted in Figure \ref{fig:t-sne_projections}, are used to support interpretation of the performance landscape by highlighting sets of configurations that exhibit similar accuracy-runtime trade-offs, and by identifying configurations that consistently underperform or behave anomalously.

\subsubsection{Friedman Test for Global Performance Differences}
\label{friedman}

Following exploratory analysis, we apply the Friedman test to formally assess whether statistically significant performance differences exist among the evaluated parameter configurations \cite{friedman37}. The Friedman test is a non-parametric alternative to repeated-measures ANOVA (ANalysis Of VAriance) and is well-suited to this study, as all configurations are evaluated on the same set of injected signals, and the performance metrics do not satisfy normality assumptions.\\

\begin{figure}[h!]
    \centering
  {%
    \includegraphics[width=\linewidth]{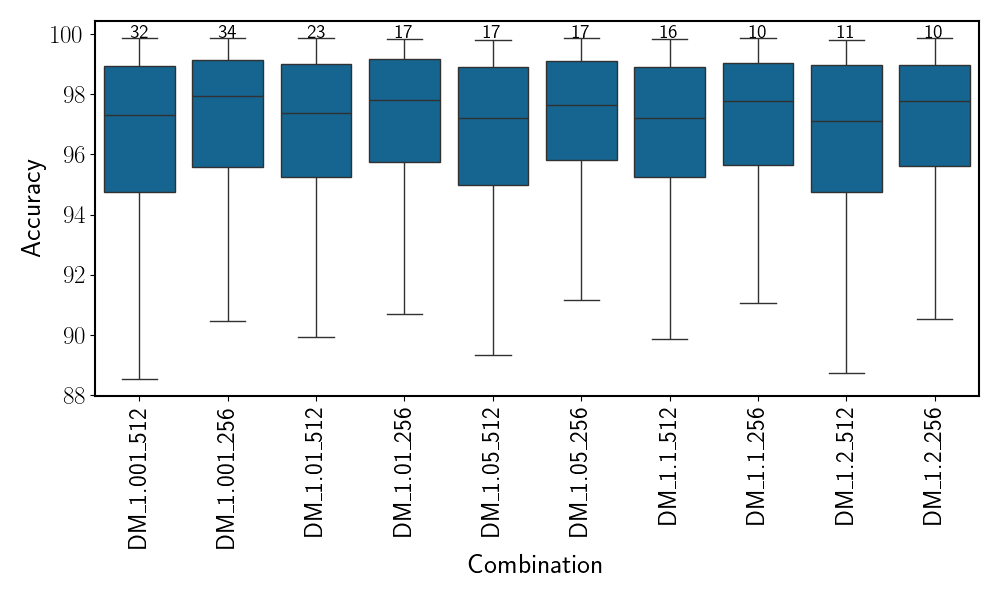}}%
  \hfill
  {%
    \includegraphics[width=\linewidth]{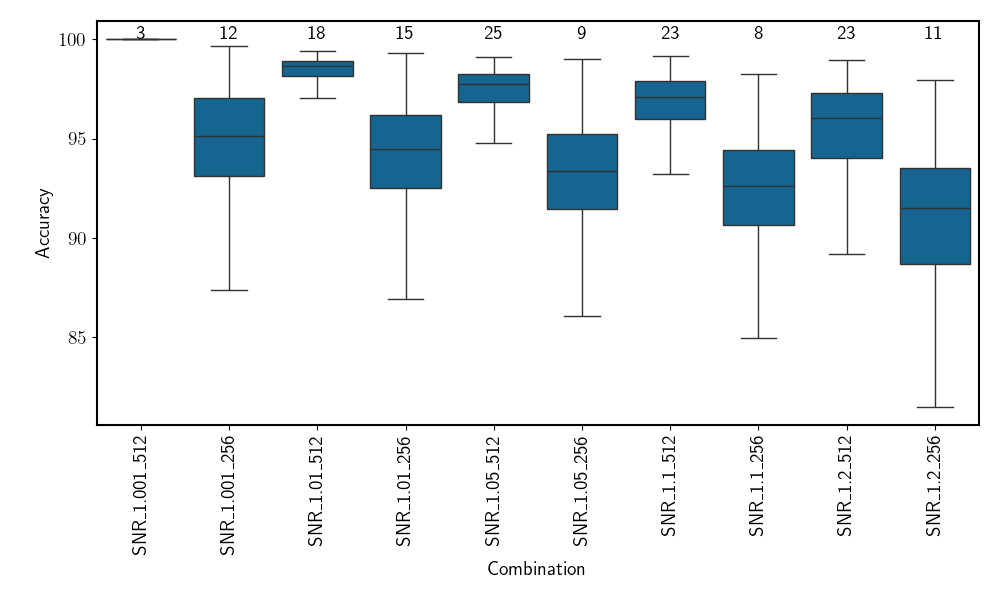}}%
        
    \caption{Boxplots of Friedman test results for DM (top) and SNR (bottom). The numerical values represent outlier counts per combination. The labels are set according to the \texttt{dm\_tol} and boxcar width parameter values.}
    \label{fig:boxplots}
\end{figure}

Each parameter configuration is ranked according to a global performance metric, and the Friedman statistic evaluates whether the observed ranking differences across configurations are greater than would be expected by chance. This rank-based approach provides a robust global test of whether parameter choice has a statistically significant effect on detection accuracy or runtime performance. The results of the best 10 configurations were represented using box plots (See Figure \ref{fig:boxplots}); the box plot results of ToA were omitted since the most influential results came from performance of DM and SNR.

\subsubsection{Nemenyi Post-hoc Pairwise Comparisons}

When the Friedman test indicates statistically significant differences, post-hoc pairwise comparisons are conducted using the Nemenyi test \cite{nemenyi63}. The Nemenyi test compares the average ranks of all pairs of configurations and determines whether their differences exceed a critical difference threshold.\\

This procedure enables the identification of specific parameter configurations that perform significantly better or worse than others across the full dataset. Results are conveniently visualised using critical difference diagrams, which group configurations that are statistically indistinguishable and highlight those that exhibit superior overall performance. Together with the cluster analysis, these results provide an interpretable basis for identifying optimal configurations.

\section{Results}
\label{sec:results}

\subsection{Detection Accuracy Across Parameter Configurations}
\label{det_accuracy_param_configs}

Detection accuracy was evaluated across the full grid of Heimdall parameter configurations using the metrics defined in Section \ref{eval_metrics}. For each configuration, recovered candidate properties were matched to injected synthetic bursts, and relative errors in DM, SNR, and ToA were computed. Performance was summarised using several statistical error metrics, including the mean absolute error (MAE), mean squared error (MSE), root mean squared error (RMSE), and mean absolute percentage error (MAPE). The results, as presented in Table \ref{tab:stat_metrics}, summarise the behaviour of each configuration across the entire dataset.\\

Across all tested configurations, detection accuracy exhibited a strong dependence on both DM tolerance and boxcar filter width. Configurations employing lower DM tolerance values consistently achieved improved DM recovery, as expected from finer sampling of the DM space. However, this improvement was not uniform across all injected DMs: at higher DMs, coarse DM grids produced noticeably larger relative DM errors, indicating increased susceptibility to dedispersion mismatch for highly dispersed signals.\\

SNR recovery showed a pronounced dependence on boxcar filter configuration. Configurations with limited maximum boxcar widths tended to underestimate the SNR of broader pulses, particularly for injected bursts with full-width at half-maximum durations exceeding several tens of milliseconds. Conversely, configurations allowing excessively large boxcar widths exhibited increased variance in SNR error for narrow pulses, reflecting the integration of additional noise when filter widths significantly exceeded the intrinsic pulse duration.\\

ToA accuracy was generally robust across most parameter configurations, with median relative timing errors remaining small compared to the intrinsic pulse widths. Nevertheless, configurations with coarse DM tolerance or mismatched boxcar ranges exhibited suggested systematic timing offsets, particularly for low-SNR bursts where imperfect dedispersion led to asymmetric pulse recovery. These effects were most pronounced for bursts near the detection threshold, highlighting the interaction between dedispersion precision and matched-filter alignment.\\

When considered jointly as was shown in Figure \ref{fig:t-sne_projections}, the three accuracy metrics reveal clear trade-offs between sensitivity and robustness. Configurations optimised for fine DM resolution improved DM and ToA accuracy but showed diminishing returns in SNR recovery relative to their increased computational cost. Conversely, configurations prioritising reduced computational complexity exhibited degraded recovery of burst properties, particularly for broad or highly dispersed signals.\\

These results demonstrate that detection accuracy cannot be optimised independently of parameter interactions. Instead, optimal performance emerges from configurations that balance dedispersion granularity with matched-filter coverage, motivating the multi-dimensional analysis and statistical comparison presented in the following sections.

\subsection{Runtime Scaling and Real-Time Performance}
\label{run_scaling}

\begin{figure*}
    \centering
    \includegraphics[width=\textwidth]{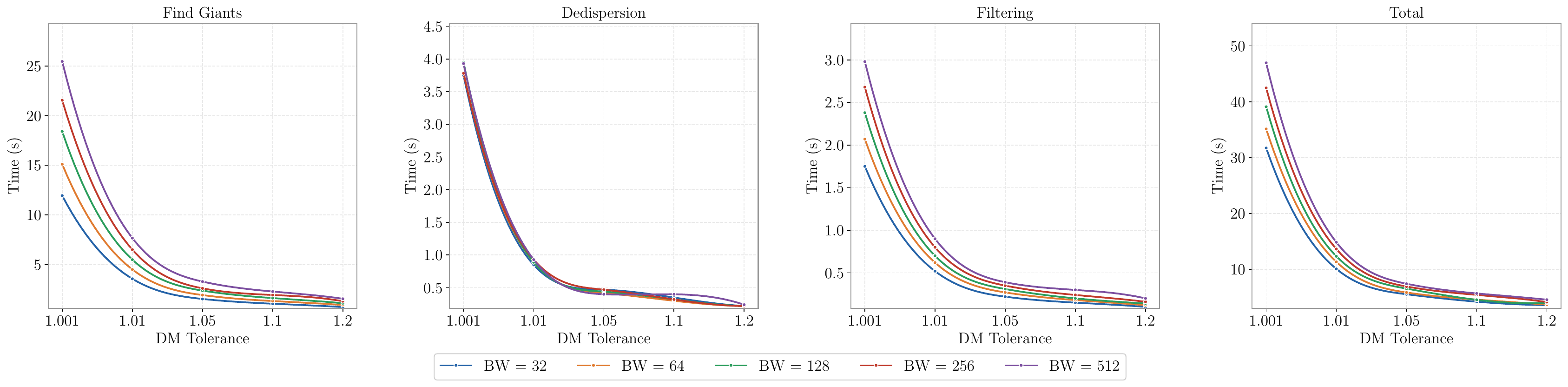}
    \caption{Execution time (s) for certain steps as a function of DM Tolerance, 
    shown for five boxcar widths (BW = 32-512). Panels show the three most time-dominant steps alongside total runtime.}
    \label{fig:performance_timings}
\end{figure*}

Runtime performance was evaluated for all parameter configurations by measuring the total processing time required to analyse each filterbank file, as described in Section \ref{run_performance}. Processing times were compared against the duration of the input data to assess real-time feasibility under different parameter choices.\\

Across the explored parameter space, runtime exhibited a strong dependence on both DM tolerance and boxcar filter configuration, as can be observed in Figure \ref{fig:performance_timings} (see also Table \ref{tab:timings_performance} for a more in depth numerical comparison). Configurations employing finer DM tolerance values incurred substantially higher computational cost due to the increased number of trial DMs. This effect was approximately linear within the tested range, reflecting the brute-force nature of incoherent dedispersion. Conversely, configurations with coarser DM tolerance reduced processing time at the expense of reduced dedispersion fidelity.\\

The maximum boxcar filter width also contributed significantly to runtime variability. Increasing the upper bound of the boxcar range increased the number of convolution operations applied to each dedispersed time series, resulting in longer processing times. While the impact of boxcar width on runtime was less pronounced than that of DM tolerance, configurations with large boxcar ranges consistently showed higher processing overhead.\\

Gulp size played a secondary but non-negligible role in determining throughput. Larger gulp sizes improved GPU utilisation by reducing kernel launch overhead and enabling more efficient memory access patterns. However, beyond a certain threshold, increasing the gulp size yielded diminishing returns, indicating that memory constraints and data transfer overheads began to dominate. In addition, large gulp sizes are not desirable in operational scenarios where low-latency triggering is required, such as saving raw voltage data or issuing alerts to external facilities for rapid follow-up observations. Larger gulps also increase memory usage, which can limit the degree of parallelisation achievable when multiple Heimdall instances are executed concurrently on the same GPU, or when Heimdall is extended to process multiple beams simultaneously. Within the tested range, a gulp size of approximately \SI{40}{\second} provided an effective balance between throughput, memory usage, and triggering latency.\\

Importantly, several parameter configurations achieved processing speeds comfortably exceeding real-time requirements. These configurations maintained processing rates significantly faster than the input data rate, leaving sufficient headroom for additional components such as candidate selection, normalisation, filtering, and alert generation.

\subsection{Statistical Comparison of Parameter Configurations}

To formally assess whether observed performance differences across parameter configurations were statistically significant, we applied the Friedman test as described in Section \ref{friedman}. The test was performed separately for each evaluation metric, treating each injected burst as a repeated measure across configurations.\\

For all detection accuracy metrics and runtime performance, the Friedman test rejected the null hypothesis of equivalent performance across configurations at the chosen significance level. This result confirms that parameter selection has a statistically significant impact on both detection fidelity and computational efficiency.\\

\begin{figure}[ht]
    \centering
    \includegraphics[width=\linewidth]{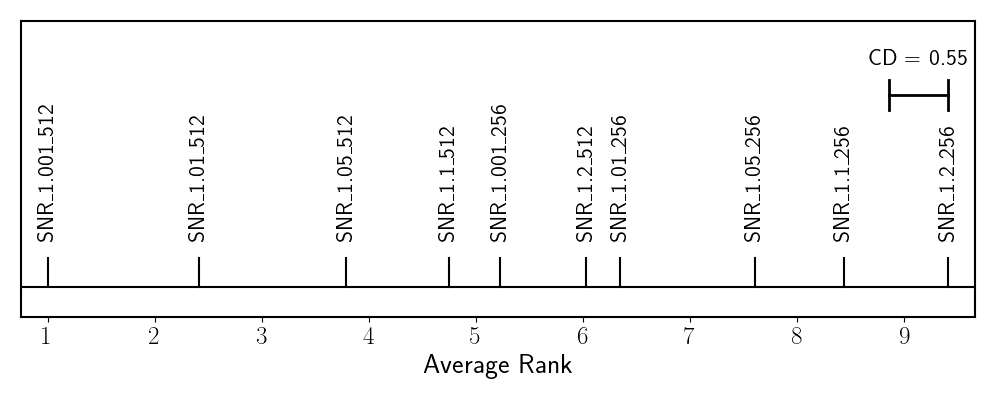}\\
    \includegraphics[width=\linewidth]{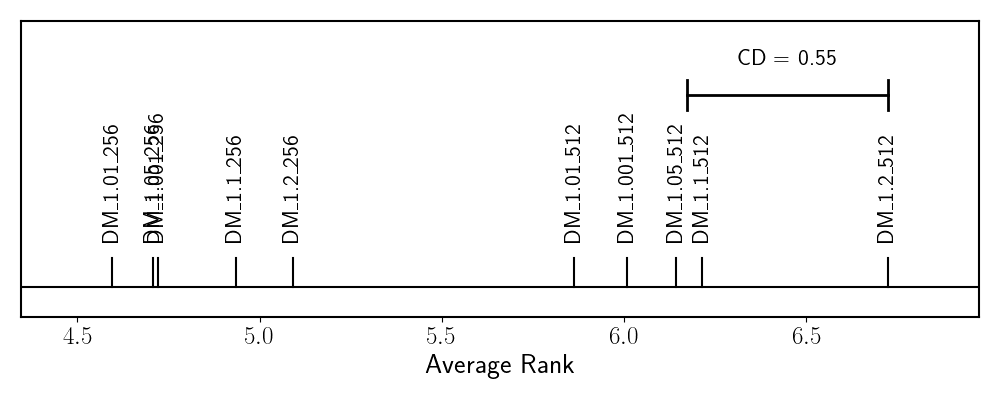}\\
    \includegraphics[width=\linewidth]{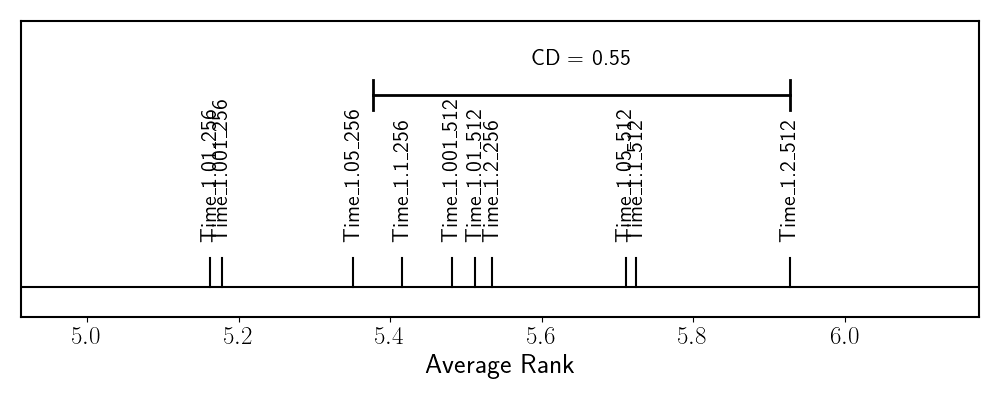}
    \caption{Critical Difference diagrams for  SNR (top), DM (middle) and ToA (bottom).}
    \label{fig:cd_diagrams}
\end{figure}

Following this global assessment, pairwise comparisons were conducted using the Nemenyi post-hoc test. Average ranks were computed for each configuration across all injected bursts, and critical difference thresholds were used to identify statistically distinguishable groups. The resulting critical difference diagrams shown in Figure \ref{fig:cd_diagrams} revealed clusters of configurations with comparable performance, as well as configurations that consistently outperformed or underperformed the rest.\\

In particular, configurations combining moderate DM tolerance with intermediate boxcar ranges achieved favourable ranks across multiple metrics, while configurations at the extremes of the parameter space such as very fine DM grids or excessively large boxcar widths, exhibited statistically significant performance degradation when accounting for both accuracy and runtime.\\

These results provide quantitative confirmation of the qualitative trends observed in Sections \ref{det_accuracy_param_configs} and \ref{run_scaling}, and support the identification of parameter configurations that achieve balanced, statistically robust performance.

\subsection{Identification of an Empirically Optimal Configuration}
\label{result}

Based on the combined evaluation of detection accuracy, runtime performance, and statistical significance, an empirically optimal parameter configuration was identified. This configuration achieved consistently strong performance across all accuracy metrics while maintaining processing speeds well above real-time requirements.\\

The selected configuration employed a DM tolerance of 1.01, a maximum boxcar width of 256 samples, and a gulp size of 40 s. Relative to other tested configurations, this setting exhibited low median errors in DM, SNR, and time-of-arrival recovery, while avoiding the substantial runtime penalties associated with finer DM grids or larger boxcar ranges.\\

Importantly, this configuration was not necessarily the best-performing option for any single metric in isolation. Rather, it provided the most favourable overall trade-off when considering all performance dimensions jointly. Statistical testing confirmed that its performance was statistically indistinguishable from the best-performing configurations for individual metrics, while being significantly more efficient than several higher-cost alternatives.\\

This result highlights the importance of multi-metric evaluation when optimising real-time transient pipelines. Parameter choices that maximise sensitivity alone may impose unnecessary computational overhead, while overly aggressive performance optimisation can degrade detection fidelity. The identified configuration represents a balanced compromise suitable for operational deployment.

\subsection{Cluster Structure in Performance Space}

To further explore relationships among parameter configurations, cluster analysis was performed using the method described in Section \ref{cluster}. Dimensionality reduction with t-SNE revealed a clear structure in the combined accuracy-runtime performance space, with configurations forming distinct groups corresponding to different trade-off regimes.\\

HDBSCAN clustering identified several stable clusters, each characterised by similar performance behaviour. One cluster comprised configurations with fine DM tolerance and large boxcar ranges, which achieved high detection accuracy but incurred substantial computational cost. Another cluster contained configurations prioritising computational efficiency, characterised by coarse DM grids and limited boxcar coverage, but exhibiting degraded recovery of injected burst properties.\\

Notably, the empirically optimal configuration identified in Section \ref{result} resided within a cluster that balanced accuracy and runtime, distinct from both the high-cost, high-sensitivity cluster and the low-cost, low-accuracy cluster. This clustering result provides additional support for the robustness of the selected configuration and demonstrates that it occupies a stable region of the performance landscape rather than representing an isolated or anomalous case.\\

Together, the clustering and statistical analyses reinforce the conclusion that systematic, data-driven evaluation of pipeline parameters can reveal structured performance regimes and guide informed optimisation decisions for real-time FRB search pipelines.

\section{Discussion}
\label{sec:discussion}

The results presented in Section \ref{sec:results} demonstrate that the performance of incoherent dedispersion-based FRB search pipelines is strongly influenced by parameter-level choices, and that these effects extend beyond simple sensitivity considerations to include computational efficiency and operational feasibility. By systematically evaluating detection accuracy and runtime across a controlled parameter space, this study provides quantitative insight into how dedispersion granularity, matched-filter coverage, and buffering strategies interact to shape overall pipeline behaviour.\\

A key outcome of this work is the identification of an empirically optimal configuration that balances detection fidelity with real-time processing requirements. Rather than maximising performance along a single metric, the selected configuration represents a compromise that achieves robust recovery of injected burst properties while maintaining substantial computational headroom. This finding underscores the importance of multi-dimensional optimisation in real-time transient searches, where sensitivity gains achieved through finer parameter sampling may be offset by disproportionate increases in computational cost.\\

The observed trade-offs between DM tolerance and detection accuracy are consistent with expectations from incoherent dedispersion theory. Finer DM grids reduce temporal smearing and improve parameter recovery, particularly at high DMs, but incur a near-linear increase in computational load due to the brute-force nature of the algorithm. Similarly, the influence of boxcar filter configuration reflects the role of matched filtering in SNR recovery: insufficient filter coverage degrades sensitivity to broad pulses, while excessively large filters increase noise integration and runtime without commensurate gains in detection fidelity. These results highlight that commonly adopted parameter choices may not be optimal when evaluated in a holistic performance framework.\\

From a computational perspective, the runtime analysis confirms that real-time processing is achievable on a single GPU for a wide range of parameter configurations, provided that buffering and filter ranges are chosen judiciously. The observed saturation of throughput gains at larger gulp sizes suggests that memory access and data transfer overheads become limiting factors beyond a certain scale, emphasising the need to consider hardware characteristics when tuning pipeline parameters. While the absolute runtime values reported here are specific to the tested hardware, the relative trends are expected to generalise to similar GPU-based implementations.\\

Although this study is motivated by observations from the Northern Cross radio telescope, the methodology and conclusions are broadly applicable to other FRB search pipelines employing incoherent dedispersion and matched filtering. The use of synthetic injections enables controlled comparison across parameter configurations and avoids confounding effects introduced by radio frequency interference (RFI) or unknown source properties. However, this approach also introduces limitations. The injected pulses adopt simplified Gaussian temporal profiles and do not capture the full complexity of real FRB signals, such as scattering tails, spectral structure, or sub-burst morphology. As a result, the absolute performance metrics reported here should be interpreted as relative indicators rather than definitive sensitivity limits.\\

Future extensions of this work could address these limitations by incorporating more realistic injection models, including scattered pulse profiles and frequency-dependent structure, as well as by validating the identified parameter configurations on real observational data. In addition, the systematic evaluation framework presented here could be extended to assess other pipeline components, such as RFI mitigation strategies or machine learning based candidate classification stages. More ambitiously, adaptive pipelines that dynamically adjust processing parameters in response to data quality or observing conditions could further improve real-time performance while preserving sensitivity.\\

Overall, this study demonstrates that systematic, data-driven evaluation of pipeline parameters can yield meaningful improvements in both detection accuracy and computational efficiency. As FRB search efforts continue to scale in data volume and complexity, such approaches will be increasingly important for ensuring that real-time transient pipelines operate at their full potential.

\section{Conclusion}
\label{sec:conclusion}

In this work, we have presented a systematic, data-driven evaluation of key processing parameters in an incoherent dedispersion based FRB search pipeline using the Heimdall single-pulse detection software. By employing a controlled synthetic injection framework, we quantitatively assessed how parameter-level choices influence both detection accuracy and computational performance under realistic observing conditions.\\

Our results demonstrate that commonly used pipeline parameters can exhibit substantial trade-offs between sensitivity and runtime, and that optimal performance cannot be achieved by maximising individual metrics in isolation. Through joint analysis of detection fidelity, processing throughput, and statistical significance, we identified an empirically optimal configuration that achieves robust recovery of injected burst properties while maintaining processing speeds comfortably exceeding real-time requirements on a single GPU. This configuration balances dedispersion granularity, matched-filter coverage, and buffering strategy, and avoids the computational overhead associated with more aggressive parameter choices.\\

Future work will focus on extending this framework to incorporate more realistic signal models, validation on real observational data, and integration with machine learning based candidate classification and adaptive pipeline strategies. As real-time radio transient searches continue to expand in scale and complexity, systematic parameter evaluation will play a key role in ensuring efficient and reliable detection performance.

\section*{Acknowledgements}

Part of the research activities described in this paper were carried out with the contribution of the NextGenerationEU funds within the National Recovery and Resilience Plan (PNRR), Mission 4 - Education and Research, Component 2 - From Research to Business (M4C2), Investment Line 3.1 - Strengthening and creation of Research Infrastructures, Project IR0000026 – Next Generation Croce del Nord

\clearpage

\appendix
\section{Tables of Results}
\label{app1}

\begin{table}[h!]
    \centering
    \begin{tabular}{clccccc}
    \hline
    Boxcar Width & Metric & \multicolumn{5}{c}{DM Tolerance}\\
    & & 1.001 & 1.01 & 1.05 & 1.1 & 1.2\\
    \multicolumn{6}{c}{}\\
    \multicolumn{2}{c}{} & $\pm 0.01$ & $\pm 0.01$ & $\pm 0.01$ & $\pm 0.01$ & $\pm 0.01$\\
    \multicolumn{6}{c}{}\\
    \multirow{0}{0pt}{32} & MAE & 17.70 & 20.02 & 19.57 & 19.46 & 20.13\\
    & MSE & 660.83 & 736.26 & 729.47 & 736.34 & 749.35\\
    & RMSE & 25.71 & 27.13 & 27.01 & 27.14 & 27.37\\
    & MAPE & 62.42 & 68.61 & 69.83 & 70.03 & 72.61\\
    & Accuracy & 37.58 & 31.39 & 30.17 & 29.97 & 27.39\\
    \multicolumn{6}{c}{}\\
    \multirow{0}{0pt}{64} & MAE & 13.41 & 13.80 & 13.90 & 13.67 & 16.77\\
    & MSE & 412.18 & 404.80 & 418.11 & 414.91 & 548.53\\
    & RMSE & 20.30 & 20.12 & 20.45 & 20.37 & 23.42\\
    & MAPE & 36.56 & 36.54 & 37.85 & 37.17 & 47.74\\
    & Accuracy & 63.44 & 63.46 & 62.15 & 62.83 & 52.26\\
    \multicolumn{6}{c}{}\\
    \multirow{0}{0pt}{128} & MAE & 8.18 & 7.09 & 7.57 & 7.48 & 8.22\\
    & MSE & 186.40 & 157.77 & 165.80 & 162.03 & 179.29\\
    & RMSE & 13.65 & 12.56 & 12.88 & 12.73 & 13.39\\
    & MAPE & 17.25 & 15.07 & 16.29 & 15.89 & 17.97\\
    & Accuracy & 82.75 & 84.93 & 83.71 & 84.11 & 82.03\\
    \multicolumn{6}{c}{}\\
    \multirow{0}{0pt}{256} & MAE & 2.11 & 2.53 & 2.86 & 2.95 & 5.25\\
    & MSE &  26.06 & 29.94 & 32.99 & 35.18 & 82.41\\
    & RMSE & 5.11 & 5.47 & 5.74 & 5.93 & 9.08\\
    & MAPE & 3.55 & 4.32 & 4.99 & 5.15 & 12.38\\
    & Accuracy & 96.45 & 95.68 & 95.01 & 94.85 & 87.62\\
    \multicolumn{6}{c}{}\\
    \multirow{0}{0pt}{512} & MAE & 0.00 & 0.41 & 0.79 & 0.48 & 1.57\\
    & MSE & 0.00 & 0.30 & 1.41 & 0.42 & 6.54\\
    & RMSE & 0.00 & 0.55 & 1.19 & 0.65 & 2.56\\
    & MAPE & 0.00 & 0.75 & 1.55 & 0.92 & 3.10\\
    & Accuracy & 100 & 99.25 & 98.45 & 99.08 & 96.90\\
    \hline
    \end{tabular}
    \caption{Sample of results of statistical metrics obtained from analyses. In this case, this table is obtained from a random file from the dataset; the results show the SNR analysis.}
    \label{tab:stat_metrics}
\end{table}

\clearpage

\begin{table}[h!]
    \vspace{-1cm}
    \centering
    \begin{tabular}{clccccc}
    \hline
    Boxcar Width & Time Execution (\unit{\second})& \multicolumn{5}{c}{DM Tolerance}\\
    & & 1.001 & 1.01 & 1.05 & 1.1 & 1.2\\
    \multicolumn{7}{c}{}\\
    \multicolumn{2}{c}{} & $\pm 0.01$ & $\pm 0.01$ & $\pm 0.01$ & $\pm 0.01$ & $\pm 0.01$\\
    \multicolumn{7}{c}{}\\
    \multirow{0}{0pt}{32} & 0-DM Cleaning & 1.78 & 1.65 & 1.81 & 1.67 & 1.79\\
    & Dedispersion & 3.75 & 0.85 & 0.47 & 0.35 & 0.22\\
    & Baselining & 2.01 & 0.60 & 0.26 & 0.18 & 0.12\\
    & Normalisation & 1.76 & 0.52 & 0.23 & 0.15 & 0.11\\
    & Filtering & 1.75 & 0.52 & 0.22 & 0.15 & 0.10\\
    & Find Giants & 11.95 & 3.57 & 1.54 & 1.06 & 0.72\\
    & Total & 31.72 & 10.09 & 5.55 & 4.27 & 3.60\\
    \multicolumn{0}{c}{}\\
    \multirow{0}{0pt}{64} & 0-DM Cleaning & 1.77 & 1.80 & 1.66 & 1.66 & 1.66\\
    & Dedispersion & 3.79 & 0.89 & 0.42 & 0.30 & 0.22\\
    & Baselining & 2.01 & 0.60 & 0.26 & 0.18 & 0.12\\
    & Normalisation & 1.76 & 0.52 & 0.23 & 0.16 & 0.11\\
    & Filtering & 2.07 & 0.62 & 0.27 & 0.18 & 0.12\\
    & Find Giants & 15.11 & 4.51 & 1.93 & 1.33 & 0.92\\
    & Total & 35.13 & 11.35 & 5.87 & 4.59 & 3.70\\
    \multicolumn{0}{c}{}\\
    \multirow{0}{0pt}{128} & 0-DM Cleaning & 1.80 & 1.64 & 1.83 & 1.67 & 1.80\\
    & Dedispersion & 3.95 & 0.89 & 0.44 & 0.32 & 0.22\\
    & Baselining & 2.01 & 0.60 & 0.26 & 0.18 & 0.12\\
    & Normalisation & 1.77 & 0.52 & 0.23 & 0.16 & 0.11\\
    & Filtering & 2.38 & 0.70 & 0.31 & 0.21 & 0.14\\
    & Find Giants & 18.41 & 5.50 & 2.38 & 1.63 & 1.11\\
    & Total & 39.11 & 12.32 & 6.56 & 4.96 & 4.04\\
    \multicolumn{0}{c}{}\\
    \multirow{0}{0pt}{256} & 0-DM Cleaning & 1.86 & 1.80 & 1.69 & 1.81 & 1.67\\
    & Dedispersion & 3.78 & 0.93 & 0.47 & 0.32 & 0.22\\
    & Baselining & 2.02 & 0.60 & 0.26 & 0.18 & 0.12\\
    & Normalisation & 1.77 & 0.52 & 0.23 & 0.16 & 0.11\\
    & Filtering & 2.68 & 0.80 & 0.35 & 0.24 & 0.16\\
    & Find Giants & 21.55 & 6.51 & 2.82 & 1.94 & 1.31\\
    & Total & 42.47 & 13.65 & 6.94 & 5.44 & 4.15\\
    \multicolumn{0}{c}{}\\
    \multirow{0}{0pt}{512} & 0-DM Cleaning & 1.77 & 1.66 & 1.83 & 1.66 & 1.81\\
    & Dedispersion & 3.93 & 0.93 & 0.40 & 0.35 & 0.24\\
    & Baselining & 1.99 & 0.60 & 0.26 & 0.18 & 0.12\\
    & Normalisation & 1.75 & 0.53 & 0.23 & 0.16 & 0.11\\
    & Filtering & 2.98 & 0.90 & 0.39 & 0.27 & 0.18\\
    & Find Giants & 25.46 & 7.67 & 3.29 & 2.29 & 1.56\\
    & Total & 46.96 & 14.86 & 7.43 & 5.70 & 4.58\\
    \hline 
    \end{tabular}
    \caption{Sample of results of performance timings (in seconds) obtained from Heimdall. This table is obtained from a random file from the dataset; the results show the timing metrics that contributed mostly to performance, for all parameter combinations.}
    \label{tab:timings_performance}
\end{table}

\clearpage






\end{document}